\documentclass[reprint]{elsarticle}   %physica A
\usepackage{amsmath}  %physica A
\journal{Physica. A}
\usepackage{fancyhdr}
\usepackage{graphicx}

\title{A random rule model of surface growth}
\author{Bernardo A. Mello\corref{cor}}
\ead{bernardo@fis.unb.br}
\address{Physics Institute, University of Brasilia, 70919-970, Brasilia, Brazil}
\date{18/09/2014}
\cortext[cor]{Email address: bernardo@fis.unb.br}

\begin{document}

\begin{frontmatter}

\begin{abstract}
Stochastic models of surface growth are usually based on randomly choosing a
substrate site to perform iterative steps, as in the etching model~\cite{Mello01}.
In this paper I modify the etching model to perform sequential, instead of random,
substrate scan. The randomicity is introduced not in the site selection but in the choice
of the rule to be followed in each site. The change
positively affects the study of dynamic and asymptotic properties, by reducing
the finite size effect and the short-time anomaly and by increasing the
saturation time. It also has computational benefits: better use of the cache
memory and the possibility of parallel implementation.\\

Published as: Physica A 419 (2015) 762–767

DOI: 10.1016/j.physa.2014.10.064
%which is not possible in models that randomly access the substrate sites.
%\PACS{81.15.Aa, 05.10.-a}
\end{abstract}


\end{frontmatter}
%\maketitle

\section{Introduction}

Stochastic simulation of surface growth plays a major role in the studies of
fractal surface dynamics. The first model of this kind, ballistic deposition,
was presented in 1959~\cite{Vold59}. A key aspect of that model is the random
choice of sites for the deposition of new atoms.

Since then, several models have been proposed combining the random site selection
with a relaxation mechanism. Among them is the Eden model~\cite{Eden61},
random deposition with surface relaxation~\cite{Family86} the restricted solid
on solid model~\cite{Kim89} and the etching model~\cite{Mello01}.

Simulations performed with those models have made important contributions to the
field, but randomly accessing the surface elements impairs the benefits from
two main advances of modern computers: cache memory and parallelism on multiple
cores of CPUs and graphics processing units (GPUs). The reasons for that are
discussed later.

In this work I propose an alternative way of inserting randomness into
surface dynamics. Instead of randomly choosing the {\em position}, the
substrate is sequentially swept and the {\em rule} followed in each site is
randomly chosen. Composed of clearly separated rules, the etching model
is particularly suitable for this approach.

Another model that introduces sequential sweeping is synchronous ballistic
deposition~\cite{Baiod88}, but in that model the rule is always the same and
the stochasticity comes from the rule being applyed or not with a given
probability.

\section{Three Versions of the Etching Model}

The etching model, which in this paper it will be called the {\em random site etching} (RSE), was introduced in 2001 to simulate the removal of atoms
in a square lattice. Simulations based on this model have been used to explore
several aspects of fractal surface dynamics~\cite{Reis03, Reis04, Reverberi05,
Reis05, Kimiagar08, Oliveira08, Tang10, Xun12, Yong-Wei12, Yu-Ying12}.

The probability of an atom being removed in the RSE is proportional to the number of
exposed faces of that atom. The higher etching probability of the more exposed atoms
may be justified either by the bigger area available to the etching agent
or by the lower number of chemical bounds between the atom and the substrate.

Despite its introduction as a model of etching, the RSE can also
be seen as a deposition process where the probability that
an atom attaches to a site is proportional to the number of bounds
that will be formed between the new atom and the ones already in the
substrate. Discussion in this paper are mostly based in this interpretation.

Whatever the interpretation, the etching model is usually implemented
with the surface moving towards positive height, as if it was
deposition from above or etching from below.
The iterative procedure of the one-dimensional version of the RSE is:
\begin{itemize}
\item randomly chooses $i \in \{1,\dots,L\}$;
\item if $h_{i+\delta} < h_i$, do $h_{i+\delta} = h_i$, with $\delta=\pm 1$;
\item $h_i = h_i+1$.
\end{itemize}
In the $d$-dimensional case, $i$ and $\delta$ are vectors and $\delta$ runs
over the $2^d$ first neighbors.  If $L$ is the substrate length along each
direction, the total number of sites is $L^d$. With the time unit defined as
the average time of one deposition at each exposed face, one iteration
corresponds to the advance of $1/L^d$ in time.

Careful analysis of the RSE algorithm revel that the deposition
of one atom is proportional to the number of the neighbors it will have.
The model I presented below keeps this property but affect only the atom
$i$, preserving the state of the neighbors.
This model will be called {\em random rule etching} (RRE) to clearly
distinguish it from the original {\em random site etching} (RSE).

While the iterative update of RSE is performed in the same data structure $h$,
the RAE require two such structures, $h^1$ and $h^0$, therefore using twice as
much memory.

In the RRE the substrate is sequentially accessed and for each site $i \in \{1,
\ldots, L\}$ of the one-dimensional surface the following steps are performed
\begin{itemize}
	\item randomly chooses $\delta \in \{-1, 0, 1\}$;
	\item if $\delta=0$ make $h_i^1 = h_i^0$ + 1;\\
	 otherwise make $h_i^1 = \max(h_i^0, h_{i+\delta}^0)$ .
\end{itemize}
After sweeping the lattice, $h_0$ is replaced by $h_1$.
In the $d$-dimensional case $\delta \in [-d,\dots,d]$. The
neighbor to appear in the $\max$ function is located along the direction
indicated by the modulus of $\delta$. There are two neighbors along each
direction, discriminated by the sign of $\delta$. Each RSE iteration performs
the $2d+1$ possible rules of
the RRE, therefore, one RSE step corresponds to $2d+1$ RRE steps. A complete RRE
substrate scan corresponds to the advance of $1/(2d+1)$ in time.

\begin{figure}
\begin{center}
\includegraphics[scale=0.69]{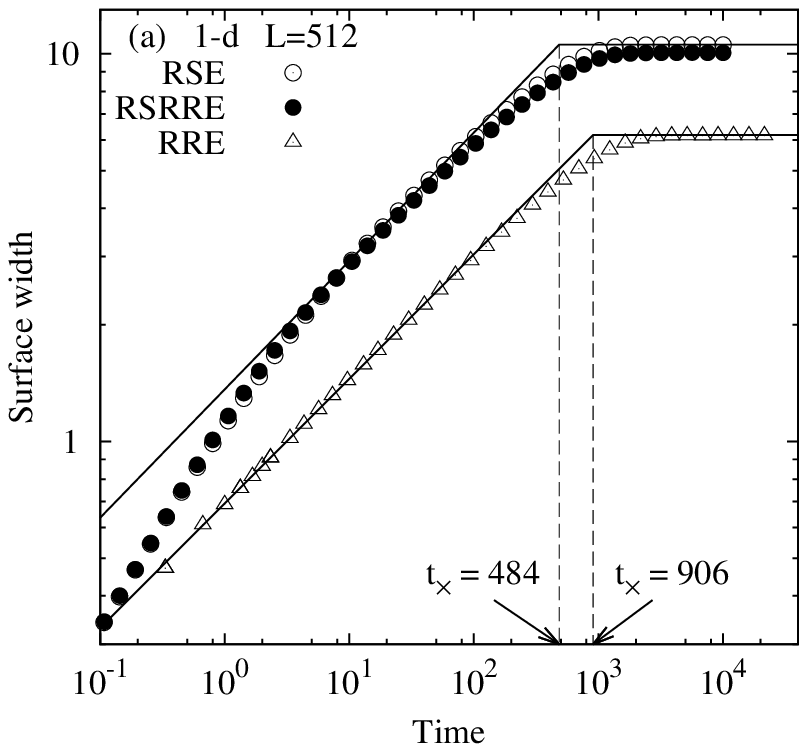} \hspace{-0.5cm}
\includegraphics[scale=0.69]{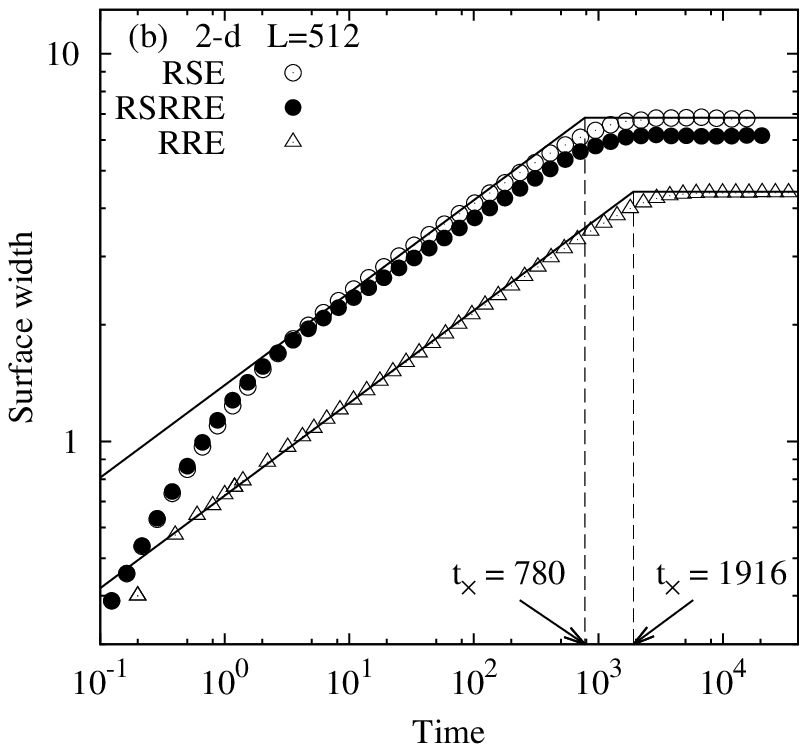}
\end{center}
\caption{Surface width as a function of time for a) one-dimensional and
b) two-dimensional substrates with $L=512$. If different values of $L$ are used,
the data points within each dimension are identical for $t\ll t_\times$.}
\label{width}
\end{figure}

For the only reason of exploring the effects of randomicity either in site
selection or in the rule selection I define the {\em random site random rule
etching} (RSRRE) model. This consists of the RRE with random, instead of
sequential, site choosing.

\section{Results and Discussion}

According to the Family-Vicsek scaling relation~\cite{Family85}, the surface width
depends on $t$ as a power law with exponent $\beta$ for $t\ll t_\times$, and
saturates to $w_s$ for $t\gg t_\times$, as depicted in Figure \ref{width}. The
values of $w_s$ and $t_\times$ scales with $L$ as, respectively, $L^\alpha$ and
$L^z$.

In Figure \ref{width} a remarkable deviation from the power law is observed in
the RSE at short times. Such anomaly is present in all models cited in the
second paragraph of this paper~\cite{Jullien85, Family86,Kim89, Mello01}. It
undermines the determination of the exponent $\beta$ and any other quantity
related to the roughening process.

The evolution of the surface width of RRE in the same figure shows a much smaller
short time anomaly, either in one or in two dimensions. Similar results occur
in higher dimensions. That result is striking since surface width $w\ll1$ at $t\sim 1$ and barely
one layer has being grown.  It is worth remembering that the substrate is swapped
$2d+1$ times at each time unit, thought that property alone cannot explain the
distinct behaviour, as discussed in the next paragraphs.

Determining the reason for the reduction in the short-time anomaly is an
inescapable question, since it will help the improvement of other models. For
that reason simulations were conducted with the RSRRE model. In Figure
\ref{width} one can observe that the surface width evolution of that model is
almost exactly the same as that of the RSE. Therefore, if the site is randomly
chosen, almost the same effects are observed with of the former celular
automata (RSE) or the one presently proposed (RSRRE), being them equivalent in
that context.

By comparing the three data sets of both graphs of Figure \ref{width} we can conclude
that the randomness in the site choosing is a major cause of the short-time
anomaly.

Exploring the conceptual differences of the RSRRE and the RRE sheds light
on the discussion, even if we fail to pinpoint which differences or how
such differences are related to the short-time anomaly.

The RRE implies one rule application at each site during the time
interval of $1/(2d+1)$. With the RSRRE a given site may be updated several times
while some other sites may get no update at all. That higher uniformity of the
sequential scan is responsible for the smaller surface width of the RRE, as
compared to the RSE and to the RSRRE.

The implementation of sequential access to the substrate imposes the use of
two data structures. Whereas each iteration of the random site corresponds to one
simulation time step, as in~\cite{Baiod88}, all application of the rule performed during one
sequential scan of the substrate happens at the same simulation time step.
Since causal connection cannot exist among these simultaneous rule applications,
information takes longer to travel the sequentially scanned substrate.

Another difference among the curves of Figure \ref{width} is the value of
$t_\times$, which is bigger for RRE. This fact, together with the minute
short-time anomaly, leads to a better fitting of the power law to the RRE data of
surface width evolution as compared to the RSE with the same substrate size.
Consequently determining RRE's $\beta$ demands smaller substrate than the RSE.
This is particularly usefull when measuring the height distribution of in the
growth region, currently an important point in the study of
KPZ~\cite{Takeuchi10, Sasamoto10, Calabrese11, Imamura12, Halpin12, Oliveira13, Alves14}, or other studies of surface dynamic~\cite{Moriconi10,Saberi08}.
To demonstrate it, I determined the exponents $\alpha$ and $\beta$ by
following the procedure proposed in \cite{Reis01}, described bellow.

\begin{figure}
\begin{center}
\includegraphics[scale=0.69]{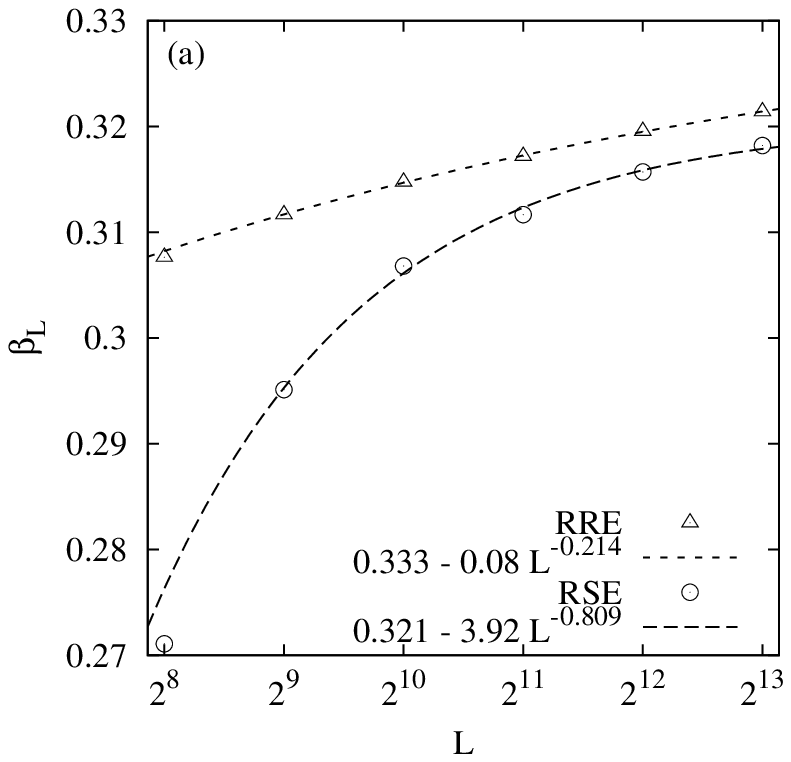} \hspace{-0.5cm}
\includegraphics[scale=0.69]{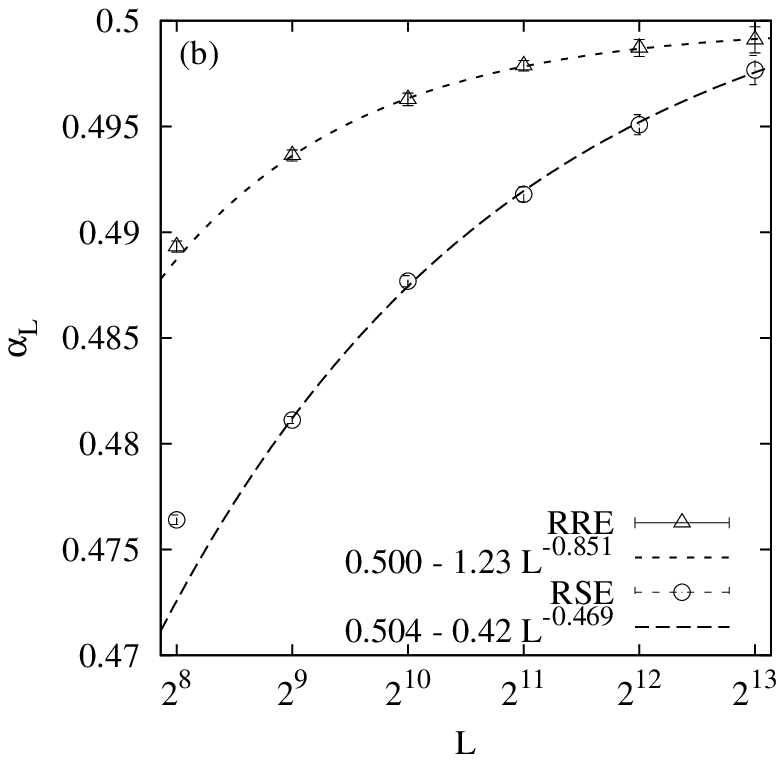}
\caption{Dynamic exponents obtained from following the procedure
described in \cite{Reis01} (see text). The fittings used
the five rightmost points of each data sate.} \label{exponents}
\end{center}
\end{figure}

For each value of $L$, linear regressions like those of figure \ref{width},
are done over the interval $[t_\text{min},t_\text{max}]$.
Considering the differences in
the short-time anomalies, I adopted $t_\text{min}=50$ for RSE
and $t_\text{min}=5$ for RRE, regardless the values of $L$.
$t_\text{max}$ is the
greatest value for which the Pearson correlation coefficient $r<0.9999$.
Figure \ref{exponents}a shows the dependence on $L$ of the resulting
regression coefficient $\beta_L$.

%\begin{figure}
%\begin{center}
%\includegraphics[scale=1]{exponents.eps}
%\end{center}
%\caption{Dynamic exponents obtained from curves like the ones
%of Figure \ref{width}. $\alpha(L) = \log ( w_s(L)/w_s(L/2))/\log(2)$ and
%$z(L) = \log ( t_\times(L)/t_\times(L/2))/\log(2)$. The dashed
%lines are the values of the parameters of the one-dimensional KPZ
%universality class, $\alpha=1/2$, $\beta=1/3$, and $z=3/2$.} \label{exponents}
%\end{figure}

The exponent $\beta$ is the limit $\beta_{L\rightarrow\infty}$,
which can be estimated from the fitting of equation
\begin{equation}
\beta_L \approx \beta + AL^{-\lambda} . \label{betaL}
\end{equation}
In Fig. \ref{exponents}a we can see the result of the fitting
to the five rightmost points of each model. While the data
points of RRE sit nicely on the curve, the same cannot be
said of the RSE points. It indicates that bigger RSE substrates are
required to obtain truthful values, as was done in \cite{Reis01},
which went up to $L=16 384$. The resulting values of $\beta$ are
0.321 and 0.333 for RSE and RRE, respectively, being 1/3 the value
expected for a model belonging to the KPZ universality class.

For each size $L$ the value of $\alpha$ was obtained as
\begin{equation}
\alpha_L = \frac{1}{\log 2}\log\frac{w_s(L)}{w_s(L/2)},
\end{equation}
where $w_s$ is the saturated width obtained at $t\gg t_\times$.
Simulations were ran until the relative error of $w_s$ be less the 0.1\%.
The value obtained are shown in Figure
\ref{exponents}b. That exponent is not affected by the short-time anomaly but by
the finite size effects~\cite{Ghaisas06}. From the
faster convergence  of the RRE as compared to RSE, for large values of $L$, we conclude
that the finite size effects are smaller in RRE than in RSE.
The asymptotic value $\alpha = \alpha_L$ is obtained by doing the
fitting to an equation equivalent to equation (\ref{betaL}),
resulting in $\alpha = 0.504$ in RSE and $\alpha = 0.500$ in RRE.

The exponents obtained in \cite{Reis01} for the RSE, using
the same procedure, are $\alpha=0.507$ and $\beta=0.339$ in $d=1$
compatible with our results. For the RRE in $d=2$, we found $\alpha=0.399$ and
$\beta=0.242$, while they found only $\alpha=0.360$, for RSE, but sugested that
the exact value should be $\alpha=0.4$. These results suggest
that both models belong to the same universality class in $d=1$ and $d=2$.
Nevertheless, although both models express the same removal probability, the
resulting dynamics are not exactly the same.

\section{Computational Benefits}

Since modern CPUs are much faster than the access rate to the main memory, a
small and fast memory, called cache, is used to keep the data which are being
read or written by the CPU. Dedicated circuits try to anticipate the
next segment of the main memory to be accessed and bring it to the cache
before it is used. If data which is not already cached is requested, it must be
fetched from the main memory, and the operation takes much longer than
when the cache contains the data.

Programs that update data sequentially leads to a memory access pattern easily
predicted by the cache managing circuit. Such applications may achieve an almost
100~\% success rate in loading data to the cache before it is requested by the
CPU.

\begin{figure}
\begin{center}
\includegraphics[scale=0.69]{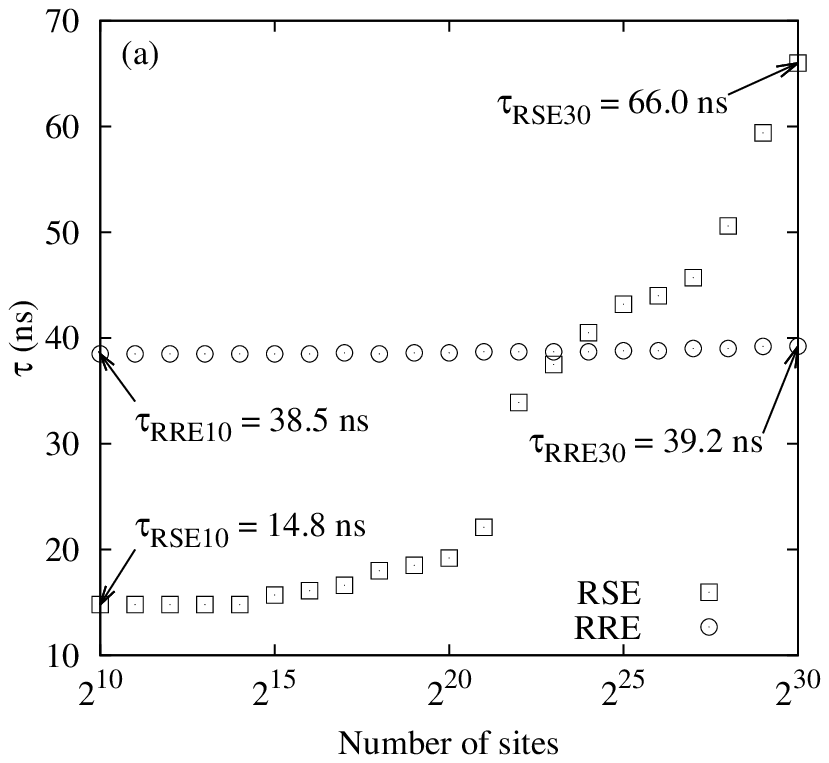} \hspace{-0.5cm}
\includegraphics[scale=0.69]{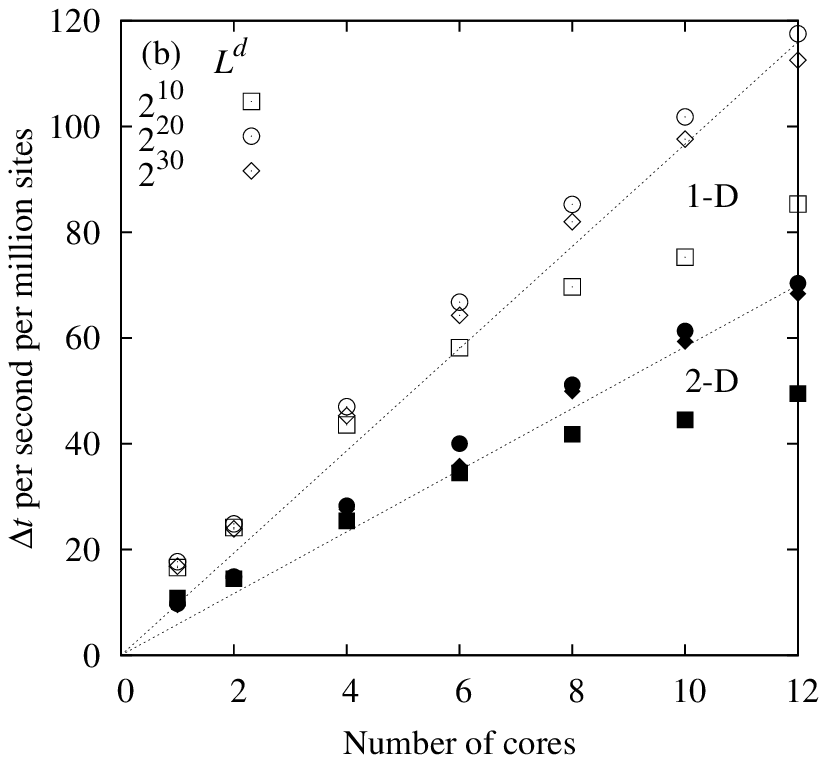}
\end{center}
\caption{(a) CPU time ($\tau$) of the stand alone simulation of one deposition of the etching
model as function of the substrate size. Each site is stored as a 2 bytes
integer in a computer with an Intel Core i5-2320 processor
(6M cache, 3 GHZ) and 6 GB of memory. Similar curves were
obtained in 2 and 3 dimensions.
With $d=2$, $\tau_{\text{RSE}10} = 24.1$ ns, $\tau_{\text{RSE}30} = 127.8$ ns,
$\tau_{\text{RRE}10} = 61.6$ ns, and $\tau_{\text{RRE}30} = 62.9$ ns.
With $d=3$, $\tau_{\text{RSE}12} = 34.6$ ns, $\tau_{\text{RSE}30} = 172.3$ ns,
$\tau_{\text{RRE}12} = 85.0$ ns, and $\tau_{\text{RRE}30} = 86.4$ ns.
(b) The vertical axis represents the advance in time of the simulation of one
million sites for one second of running. Hollow figures, one dimension, solid
figures, two dimensions. Remember that each site is iterated $2d+1$ times at
each time unit. Machine with two Intel Xeon X650 processors,
each with 6 cores. The lines are only guides to the eyes.}
\label{cache}
\end{figure}

The random choice of the deposition site present in most stochastic models of
surface dynamics makes it impossible to anticipate which part of the memory
will be accessed in the next iteration. This is not a serious problem when the
whole substrate can be stored in the cache, however, it drastically degrades
performance for bigger substrates. Computation time steps up when the used
memory $2 N_\text{sites} \approx 6 \text{MB}$, i.e., when $N_\text{sites}
\approx 10^{22}$, as shown in Figure \ref{cache}$a$. Complete understanding of how
used memory affects the computation time requires the discussion of other
optimization mechanisms, which is beyond the scope of this paper.

Figure \ref{cache}$a$ shows that the simulation time is almost constant for the RRE model,
where the surface is sequentially scanned.

Another caveat of the random site selection is that the workload cannot be
distributed among several processing units, each of them responsible for one
sector of the substrate, as can be done in sequential access algorithms. This
restriction prevents the use of computer parallelism in the evolution of one
substrate. The limitation can be circumvented in multi-core computers by running
independent substrates in each core, but the strategy cannot be used if the
substrate size is comparable to the machine memory. In this case processing
power will be wasted. Such situations occur when simulating high
dimensionality substrates.

That drawback was made worse by the popularization of the use of graphic
processing units (GPU) for scientific calculations, whose processing power may
be more than a hundred times higher than that of desktop
computers~\cite{Sanders10}. The gain is based on massive parallelism and GPUs
may have hundreds or thousands of processing units with just a few gigabytes of
memory. The ratio between processing power and memory size is much higher in
the GPUs than in the conventional computers. This puts a more stringent limit on
parallelism via the simulation of independent substrates.

Besides the sequential substrate scan, one important aspect of the RRE is that
only the site $i$ can be altered at each step, which makes code parallelization
a simple task.

Once the RRE algorithm has being implemented, it can easily be parallelized
to use the multiple cores of modern computers. One way of doing
this in C or Fortran programs is by using openMP~\cite{openMP13}. Each substrate scan is
divided in $N_T$ segments attributed to the $N_T$ available threads. To improve
efficiency, $N_T$ independent\footnote{By independent I mean having its own
internal variables.} random generators must be created, each one used
exclusively by a single thread. Care must be taken to avoid false sharing of the
random generators and other variables private to the threads~\cite{Chapman08}.
If properly implemented, good scalability is achieved for big substrates, as can be seen in
figure \ref{cache}$b$.

\section{Conclusions}

In conclusion, adapting a stochastic surface growth model to access the substrate
sites sequentially, instead of randomly, can bring several advantages:
\begin{itemize}
\item Reduction of the short-time anomalies.
\item Reduction of finite length effects.
\item Increase of $t_\times$ (useful when studying the roughening process,
	but an inconvenience when studying the steady state.)
\item Smaller substrates are required when estimating macroscopic properties.
\item Efficient use of the cache memory.
\item Parallelizable algorithm.
\end{itemize}

It was shown that the reduction in the short-time anomalies is a consequence
of changing the site selection from random to sequential. It cannot
be explained by other differences between the models, like the change in the
algorithm or the unfolding of one iteration of the RSE in to $2d+1$
iteration of RRE and RSRRE. A possible explanation is the uniform surface
scanning in the new model, while in the original model, a give site may, for
example, not be updated for several units of time.

Not all surface growth models can be converted to perform sequential scanning. The etching
model had to be altered to have only one modifiable site
at each algorithm step, otherwise, the sweeping order will affect the
dynamics. Furthermore, sequentializing is not possible if the rule applied in each cell is uniquely
determined by the surface state, i.e., the rule must have some randomicity.
The surface width evolution was barely affected
by that change, but was significantly altered by changing
the order, from random to sequential, by which the substrate sites
were accessed.

Two undesired aspects of the RRE as compared to the RSE
are the use of twice
as much memory and the division of the original step
in $2d+1$ steps which must be performed to achieve the same time evolution.
However the changes pay off in light of the benefits: increased processing
speed for big substrates and reduced short-time anomalies and
finite size effect in small and big substrates.

%\bibliography{RRE.bib}
%\end{document}

\end{document}